\pdfoutput=1
\documentclass[5p,fleqn]{elsarticle}
\usepackage[latin1]{inputenc}
\usepackage{graphicx}
\usepackage{hyperref}
\usepackage{upgreek}
\usepackage{lineno}

\begin{document}

\title{The single photon sensitivity of the Adaptive Gain Integrating Pixel Detector}

\author[hh]{Julian Becker\corref{cor1}}
\ead{Julian.Becker@desy.de}
\author[psi]{Dominic Greiffenberg}
\author[hh]{Ulrich Trunk}
\author[psi]{Xintian Shi}
\author[psi]{Roberto Dinapoli}
\author[psi]{Aldo Mozzanica}
\author[psi]{Beat Henrich}
\author[psi]{Bernd Schmitt}
\author[hh]{Heinz Graafsma}
\cortext[cor1]{Corresponding author}
\address[hh]{Deutsches Elektronen-Synchrotron,\\ Notkestr. 85, 22607 Hamburg}
\address[psi]{Paul Scherrer Institute, \\ 5232 Villigen PSI, Switzerland}

\begin{abstract}

Single photon sensitivity is an important property of certain detection systems. This work investigated the single photon sensitivity of the Adaptive Gain Integrating Pixel Detector (AGIPD) and its dependence on possible detector noise values. Due to special requirements at the European X-ray Free Electron Laser (XFEL) the AGIPD finds the number of photons absorbed in each pixel by integrating the total signal. Photon counting is done off line on a thresholded data set.

It was shown that AGIPD will be sensitive to single photons of 8~keV energy or more (detection efficiency $\gg$ 50\%, less than 1 count due to noise per 10$^6$ pixels). Should the final noise be at the lower end of the possible range (200 - 400 electrons) single photon sensitivity can also be achieved at 5~keV beam energy.

It was shown that charge summing schemes are beneficial when the noise is sufficiently low. The total detection rate of events is increased and the probability to count a single event multiple times in adjacent pixels is reduced by a factor of up to 40. 

The entry window of AGIPD allows 3~keV photons to reach the sensitive volume with approximately 70\% probability. Therefore the low energy performance of AGIPD was explored, finding a maximum noise floor below 0.035 hits/pixel/frame at 3~keV beam energy. Depending on the noise level and selected threshold this value can be reduced by a factor of approximately 10. Even though single photon sensitivity, as defined in this work, is not given, imaging at this energy is still possible, allowing Poisson noise limited performance for signals significantly above the noise floor.

\end{abstract}


\maketitle

\linenumbers

\section{Introduction}

The European X-Ray Free Electron Laser (XFEL) \cite{XFEL, TN2011-001} will provide ultra short, highly coherent X-ray pulses which will revolutionize scientific experiments in a variety of disciplines spanning physics, chemistry, materials science, and biology. 

One of the differences between the European XFEL and other free electron laser sources is the high pulse repetition frequency of 4.5~MHz. The European XFEL will provide pulse trains, consisting of up to 2700 pulses separated by 220~ns (600~$\upmu$s in total) followed by an idle time of 99.4~ms, resulting in a supercycle of 10~Hz and 27000 pulses per second.

Dedicated fast 2D detectors are being developed, one of which is the Adaptive Gain Integrating Pixel Detector (AGIPD) \cite{AGIPD1, AGIPD2, AGIPD3}. This development is a collaboration between DESY, the University of Hamburg, the University of Bonn (all in Germany) and the Paul Scherrer Institute (PSI) in Switzerland.

Many experiments at the European XFEL will require single photon sensitivity, which is an important requirement of the detection system. Due to the large variety of experiments foreseen at the European XFEL one cannot define a unique requirement for all experiments, but rather one has to optimize performance parameters like the number of noise counts/pixel/frame (false positive rate) and single photon detection efficiency (true positive rate) with respect to each individual experiment.

In this study, single photon sensitivity of AGIPD is defined as a false positive rate is below 10$^{-6}$ (less than one hit due to noise per frame of 1 megapixel\footnote{1 megapixel (MP) is 1024 $\times$ 1024 $\approx 10^6$ pixels.}) while simultaneously having an average true positive rate of more than 50\%. The single photon sensitivity is investigated for the design energy of 12.4~keV and the optional lower operating energies of 8, 5 and 3~keV \cite{cdr_spb, cdr_mid}. 

\section{The Adaptive Gain Integrating Pixel Detector (AGIPD)}

\begin{figure}[tb!]
  \includegraphics[width=0.45\textwidth]{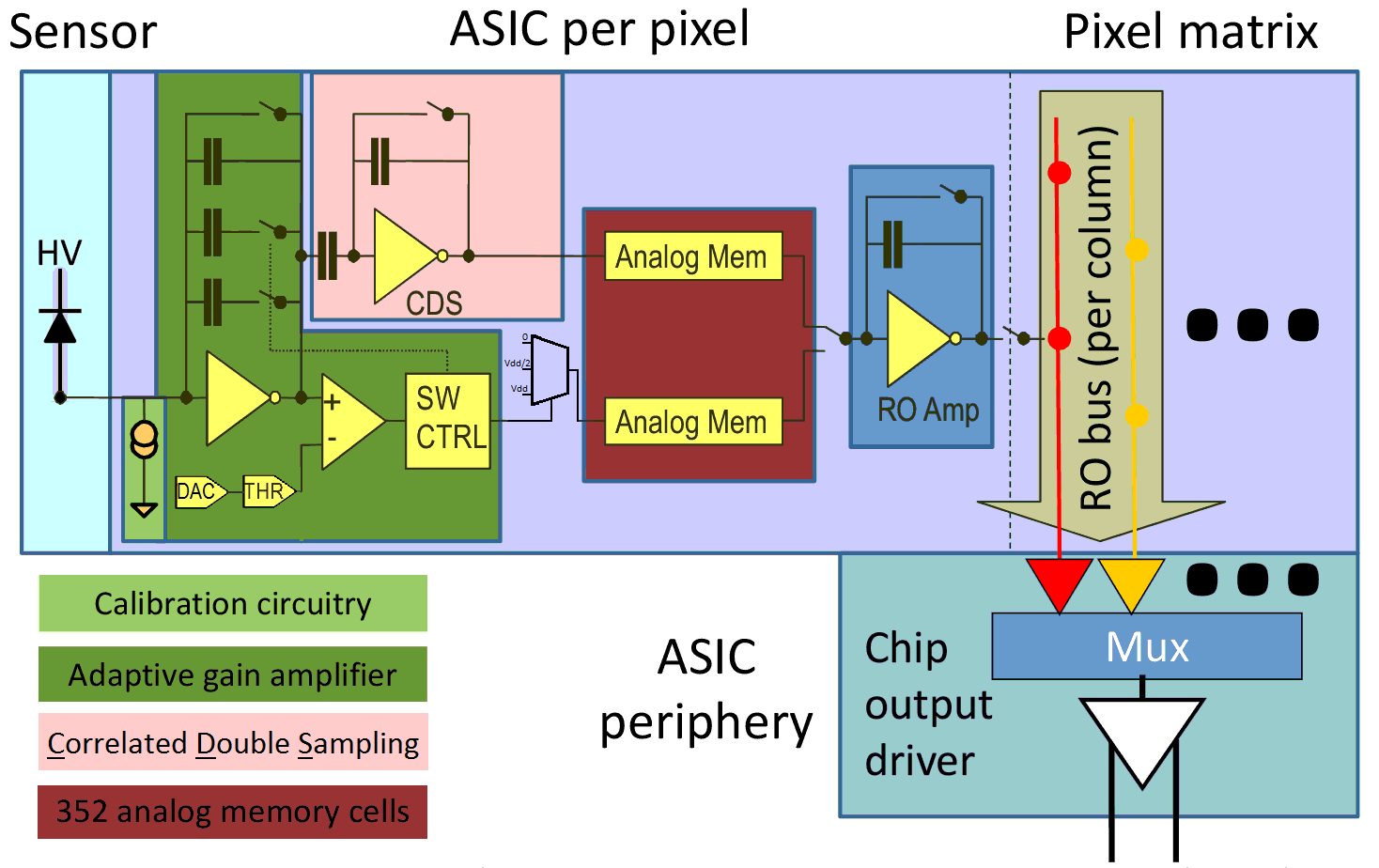}
  \centering
  \caption{Block diagram of the integrated circuitry of the AGIPD04 test-chip. A similar circuit with minor modifications will be used on the 64x64 pixel chip AGIPD10.}
  \label{block}
\end{figure}

AGIPD is based on the hybrid pixel technology. The current design goals of the newly developed Application Specific Integrated Circuit (ASIC) with dynamic gain switching amplifier in each pixel are (for each pixel) a dynamic range of more than 10$^4$ 12.4~keV photons in the lowest gain, single photon sensitivity in the highest gain, and operation at 4.5~MHz speed. An external veto signal can be provided to maximize the number of useful images by overwriting any image previously recorded during the pulse train. 

Due to the special pulse structure of the European XFEL, it is necessary to store the acquired images inside the pixel logic during the pulse train. A compromise had to be found between storing many images, requiring a large pixel area, and high spatial resolution, requiring small pixels sizes \cite{AGIPD1, AGIPD2}. The image data is read out and digitized in the 99.4~ms between pulse trains.

The AGIPD will feature a pixel size of (200~$\upmu$m)$^2$, which is sufficient to house an analog memory capable of storing 352 images. For most experiments using particle injection mechanisms the memory of AGIPD should be sufficient, as particle hit rates are usually below 10\% (see \cite{cdr_spb} and references therein). The impact of the limited number of storage cell on X-ray Photon Correlation Spectroscopy (XPCS) as intended to be used on the MID station \cite{cdr_mid} depends on the properties of the sample and has been investigated elsewhere \cite{xpcs_sampling}.

In order to provide a sufficiently high quantum efficiency at higher photon energies a silicon sensor with a thickness of 500~$\upmu$m will be used.

\subsection{Signal processing chain}
A block diagram of the full scale chip is shown in figure \ref{block}. The typical signal path comprises charge generation and transport within the sensor, charge collection and (amplified) charge to voltage conversion in the adaptive gain amplifier, Correlated Double Sampling (CDS) of the amplifier output voltage by the CDS stage, signal storage in the analog memory cells, readout of the storage cells and communication of the signal to the outside world via differential and LVDS lines. 

For pristine\footnote{After heavy irradiation and/or at elevated temperatures the signal droop of the storage cells becomes non negligible. Reducing the operating temperature has been shown to mitigate this effect \cite{droop}, detailed investigations are currently ongoing.} chips under standard operation conditions\footnote{Standard operating conditions employ about 100~ns integration time. For such short exposure times the contribution of leakage currents, even when elevated after irradiation, is negligible. Integrating over multiple pulses has been suggested for certain experiments, but the impact of this is beyond the scope of this study.} the amplifier, CDS stage and readout contribute about equally to the total noise, which for any given sample follows a Gaussian distribution\footnote{Although all observations support it, this assumption is not trivial, especially for tails more than 5 sigma away from the mean (i.e. very rare events).}. Therefore it is convenient to express the noise in terms of the Equivalent Noise Charge (ENC), which is the standard deviation (sigma) of the distribution. A more detailed noise analysis has been presented before \cite{AGIPD2}; for this study it is sufficient to simplify matters and only use the ENC as the relevant noise parameter.

\subsection{Estimated performance of AGIPD}

\subsubsection{Noise measurements}
Measurements of AGIPD02 showed an overall ENC of approximately 320 electrons \cite{iworid}. This omits the storage cell noise and the dominant remaining noise contributions originate in the Correlated Double Sampling (CDS) stage and the readout buffer. These parts of the ASIC circuitry have been modified to have a lower noise contribution and were manufactured on other test chips (AGIPD03 and AGIPD04). Measurements of these test chips are currently being performed, but not yet available.

Circuit simulations of the improved parts allow an estimation of the total noise to be around 300 electrons for AGIPD03, and a reduced noise was verified by preliminary measurements of the chip \cite{poster}. The AGIPD04 test chip features pixel rows with an increased gain in the first gain stage, which might further reduce the total noise\footnote{As CDS stage and readout buffer insert their noise 'downstream' of the preamplifier an increase in gain reduces the equivalent noise charge on the input.}.

Assuming that these results and expectations are transferable to AGIPD10, a full scale chip featuring 64~$\times$~64 pixels, one can estimate the ENC of the system to be between 200 and 400 electrons. Selected simulations with 260 and 360 electrons ENC will be presented in more detail later in this study.

\subsubsection{Quantum efficiency}

\begin{table*}[tb]
	\centering
		\begin{tabular}{c|c|c|c}
			Window type	& Aluminum [$\upmu$m]	& n$^+$ implantation [$\upmu$m]	& sensitive thickness [$\upmu$m] \\
				\hline 
			standard 		& 0.5 								& 1.2														& 497.6 \\
			thin 				& 0.1 								& 0.1														& 498.7 
		\end{tabular}
	\caption{Entrance windows investigated in this study, neglecting thin SiO$_2$ layers. The thickness of the sensor is 500 $\upmu$m, where the last 1.2~$\upmu$m are assumed to be insensitive due to the p$^+$ implantation. }
	\label{window}
\end{table*}

The quantum efficiency $QE$ can be approximated as the probability of a photon to be absorbed\footnote{All derivations in this paragraph use the cross section for photoabsorption. As a consequence 2nd order effects including in-sensor scattering are neglected.} in the sensitive volume of the sensor as a function of photon energy $E$ and angle $\theta$ between the incoming photon and the beam axis:

\begin{eqnarray}
	QE(E,\theta) &=& P_1(E,\theta) * P_2(E,\theta) ,\; \;\;\mathrm{with}\\
	P_1(E,\theta) &=&  \prod_{i} e^{-\frac{\mu_i(E) d_i}{\cos \theta}}  \; \;\;\mathrm{and} \\
	P_2(E,\theta) &=& 1-e^{-\frac{\mu_{sens}(E) d_{sens}}{\cos \theta}}.  \label{pabs}
\end{eqnarray}

$P_1$ is the product of the transmission probabilities of all $i$ materials of the entry window. It denotes the probability of a photon passing through the entry window without interacting. $\mu_i$ and $d_i$ are the linear attenuation coefficient and the thickness of material $i$ taken from \cite{nist}. For the chosen sensor thickness $P_1$ is the dominating term at energies below approximately 8~keV.

$P_2$ denotes the probability of a photon being absorbed in the sensitive volume (dominating at energies above approximately 8~keV), with $\mu_{sens}$ and $d_{sens}$ being the linear attenuation coefficient and the thickness of the sensitive material (silicon).

This approximation neglects any effects due to the finite lateral extent of the sensor material and any disturbances along the photon's flight path, especially all windows of vacuum vessels and flight tubes.

The effect of a thin (few 100~nm thickness) SiO$_2$ layer between aluminum and n$^+$ implantation is negligible\footnote{The purpose of the thin oxide layer is to increase the adhesion of the aluminum and to reduce the probability of 'spikes' of 'hillocks'. Electric contact between the implant and the metal is ensured by a sufficient number of small 'vias' in the oxide. For more information see \cite{layer1, layer2} and references therein.}. Details on the pixel layout and junction depth on the readout side can be found in literature \cite{schwandt}. Reducing the thickness of the entry window materials has been suggested as a possibility to increase the sensitivity of AGIPD to low energy photons. A comparison of the material thicknesses of a standard window and a possible thin window is listed in table \ref{window}.

\begin{figure}[tb!]
  \includegraphics[width=0.45\textwidth]{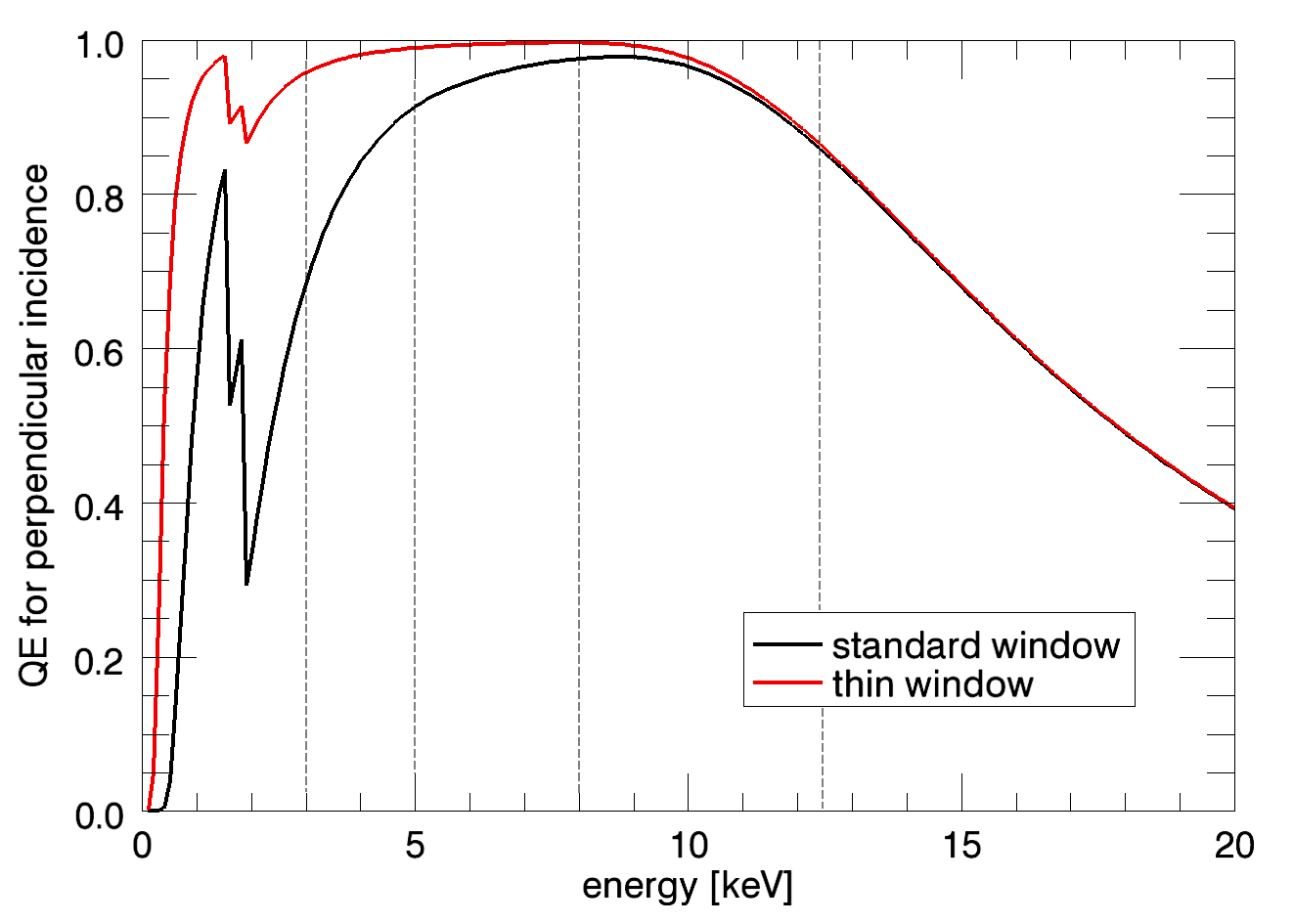}
  \centering
  \caption{Approximate quantum efficiency as a function of energy for the two different window designs. The thin window significantly improves the $QE$ for energies below approximately 5~keV. The vertical dashed lines indicate the energies investigated in this study.} \label{qe_window}
\end{figure}

The results for the quantum efficiency for both windows as a function of energy are shown in figure \ref{qe_window}. A thin window would increase the $QE$ compared to the standard window by about 40\%, 7\%, 2\% and 1\% at an energy of 3, 5, 8 and 12.4~keV, respectively.

\subsection{Off-chip charge summing algorithm}
\label{summing}
The optional off-chip charge summing algorithm used in this study is summing the response of two or more pixels to recombine energy depositions which were spread over adjacent pixels, thereby improving the sensitivity to single photons and reducing the probability to count a single photon multiple times.

As pixels are summed up, so is their noise. Assuming the noise of all pixels to be uncorrelated and to have an identical rms value of $ENC$, the $N$ summed pixels can be considered one larger pixel with a noise of $\sqrt{N}\;ENC$. It should be noted that charge summing, in contrast to pixel binning, does not reduce the spatial resolution.

In the present study pixels were not summed unconditionally but had to fulfill certain criteria:
\begin{itemize}
\item Pixels were considered suitable for summing when their energy deposition exceeded 570 electrons\footnote{The seed criterion reduces the number of possible pixel summing combinations to test, thereby increasing the overall speed. As long as it is below half (2 pixel summing) or one quarter (4 pixel summing) of the beam energy it has no influence on the result.} (approximately 2~keV), but was below threshold (seed criterion).
\item 2 pixel summing: Pixels fulfilling the seed criterion were  considered hit, when the energy summed with one of the four neighbors sharing an edge was above threshold. The hit was assigned to the pixel with the largest energy contribution (neighbor criterion). If more then one possibility to fulfill the neighbor criterion was found, the pixels with the highest total energy deposition were summed. Pixels could only be summed once.
\item 4 pixel summing: Pixels fulfilling the seed criterion and the neighbor criterion extended to 2 by 2 pixel regions (Quad criterion).
\end{itemize}

As the pixels were not summed unconditionally, the introduced selection bias increases the probability of finding a false positive beyond the rate that would be expected with $\sqrt{2}\; ENC$ noise for unconditional 2 pixel summing\footnote{An example is a pixel being just above threshold with all its neighbors being negative enough for the sum to be below threshold. Unconditional summing will not detect a hit, while the employed summing algorithm will.}. On the other hand the introduced selection bias sometimes decreased the probability of finding a false positive below the expected $2 \; ENC$ for unconditional 4 pixel summing, as summing two pixels fulfilled the neighbor criterion for most pixels.

It should be noted that charge summing is a data processing step which is done completely offline, i.e. not within the ASIC, as done e.g. in the Medipix3 chip\cite{medipix3}, or readout chain\footnote{It would be possible to include the data processing at some point in the readout chain, but this is not foreseen at the moment.}.

\section{HORUS software and comparison to measurements}

The detector response of AGIPD was calculated using the HORUS software described in \cite{AGIPD3, horus1}. HORUS has already been successfully used to simulate the performance of the Medipix3 chip \cite{david1, david2} and recently for AGIPD \cite{xpcs_sampling, horus3, last_paper}. For this study HORUS has been upgraded to include the simulation of the K-edge fluorescence of silicon, elastic (Thomson or Rayleigh) scattering and inelastic (Compton) scattering within the sensor material. 

\subsection{Comparison of measurements with simulation results}

In order to validate and verifiy the HORUS package a set of measurements using an assembly of AGIPD02 was compared to corresponding simulations. The measurements were taken using an X-ray tube and different operating conditions than those to be used for XFEL experiments. Details are presented in the folowing paragraphs. It should be noted that as the operating conditions differ the important result of the comparison is the degree of agreement between simulation and measurement, not the absolute values of the parameters used in the simulation.

\subsubsection{Measurement setup and data taking conditions}
The measurements were performed using a copper anode X-ray tube\footnote{ISO DEBYEFLEX 3003 from GE Measurement and Control Solutions, set to 50~kV and 39~mA} illuminating a high purity\footnote{99.9\% from Goodfellow GmbH, 61213 Bad Nauheim, Germany} molybdenum foil. The detector assembly was positioned at 90$^\circ$ angle with respect to the X-ray beam from the tube (the foil was positioned at 45$^\circ$ with repect to each) in order to minimize the elastically scattered photons from the x-ray beam\footnote{Although this setup eliminates most of the tube spectrum due to the low scattering probability, the intensity of the copper K-edge fluorescence photons is high enough that some of them can be detected at the sensor location.}. The distance between foil and assembly was about 15~cm, so effects of non-perpendicular photon incidence were minimal. 

The detector assembly was mounted in a custom made chip tester box that handled all communication with the ASIC. In contrast to the sensor thickness intended for the full scale detector, the ASIC was bump bonded to a silicon sensor of 320~$\upmu$m thickness having a depletion voltage of  approximately 50~V. To (over-)deplete the sensor a bias voltage of 120~V was applied. All measurements were performed without cooling at room temperature. Due to the power dissipation of the chip tester box and the ASIC the assembly temperature stabilized around 45$^\circ$ C.

In order to increase the probability to detect a photon an integration time of 1~$\upmu$s was used. For this integration time the leakage current of the sensor is no longer negligible and a noise between 350 and 400 electrons was observed. This is higher than the value of 318 $\pm$ 18 electrons previously reported for 100~ns integration time \cite{iworid}. A more detailed description of the operating conditions and their influence on the measurement results can be found in literature \cite{iworid}. A total of 5~$\times$~10$^5$ frames were acquired. For comparison with the simulations only events of a single pixel were evaluated\footnote{In this way pixel to pixel variations are excluded from the comparison.}.

In the end the measurement data used for the comparison consisted of 5~$\times$~10$^5$ statistically independent samples which were pedestal corrected\footnote{The pedestal correction ensures that the most probable measurement point is located at 0 units by offset subtraction.} and histogrammed for the comparison with a bin size of 2 AD (Analog to Digital converter) units.

\subsubsection{Simulation setup}

The HORUS simulations calculated 5~$\times$~10$^5$ pixels\footnote{As the simulations did not account for edge effects the system is ergodic. This means that the simulation of $N$ pixels  in $M$ frames produces statistically equivalent results for any combination of $N$ and $M$ as long as the product of $N$ and $M$ stays constant.}, with an average probability (Poisson distributed) to have molybdenum and copper fluorescence photons impinge on a pixel of 2.22\% and 0.102\% (equally distributed over each pixels surface). The fluorescence photons included $K_\alpha$ and $K_\beta$ lines. The probabilities of fluorescence photons as well as fine adjustments of the simulation parameters were determined from a manual optimization process to produce the best agreement of simulation and measurement. The total equivalent noise charge (ENC) at the input of the simulated system was 376 electrons, in order to account for the mode of operation mentioned above. 

In the end the simulated data used for the comparison consisted of 5~$\times$~10$^5$ statistically independent samples which were pedestal corrected and histogrammed for the comparison with a bin size of 2 AD units.

To show the effects of improper parameter tuning a second 'wrong' data set was calculated, which wrongly assumed a noise of 350 electrons, no copper fluorescence photons and no $K_\beta$ line of molybdenum.

\subsubsection{Comparison of measurement and simulation}

\begin{figure}[tb!]
  \includegraphics[width=0.45\textwidth]{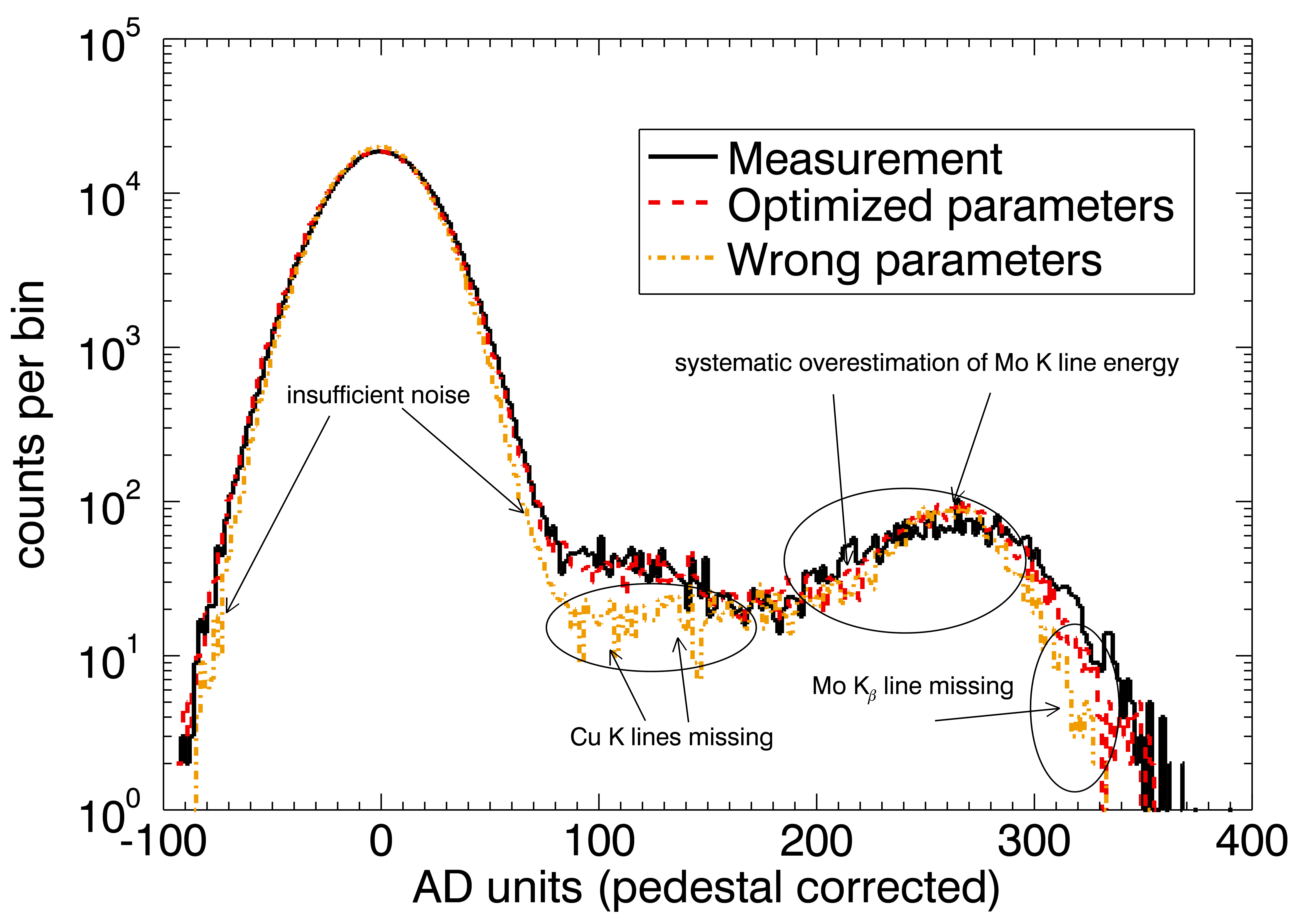}
  \centering
  \caption{Comparison of measurements performed using an AGIPD02 assembly with sensor and simulations with HORUS. The annotations have been inserted to guide the eyes. Details on the parameters used for the simulations are presented in the main text.}
  \label{direct_comp}
\end{figure}

Figure \ref{direct_comp} shows a direct comparison of the measured and simulated counts as a function of AD units using a semi logarithmic scale for better visibility of the small number of photon events. The 'wrong' data set is included for comparison. The data evaluation of experiments at the European XFEL will most probably not use the raw data in AD units but integer numbers of photons. The number of photons will be determined by thresholding. 

\begin{figure}[tb!]
  \includegraphics[width=0.45\textwidth]{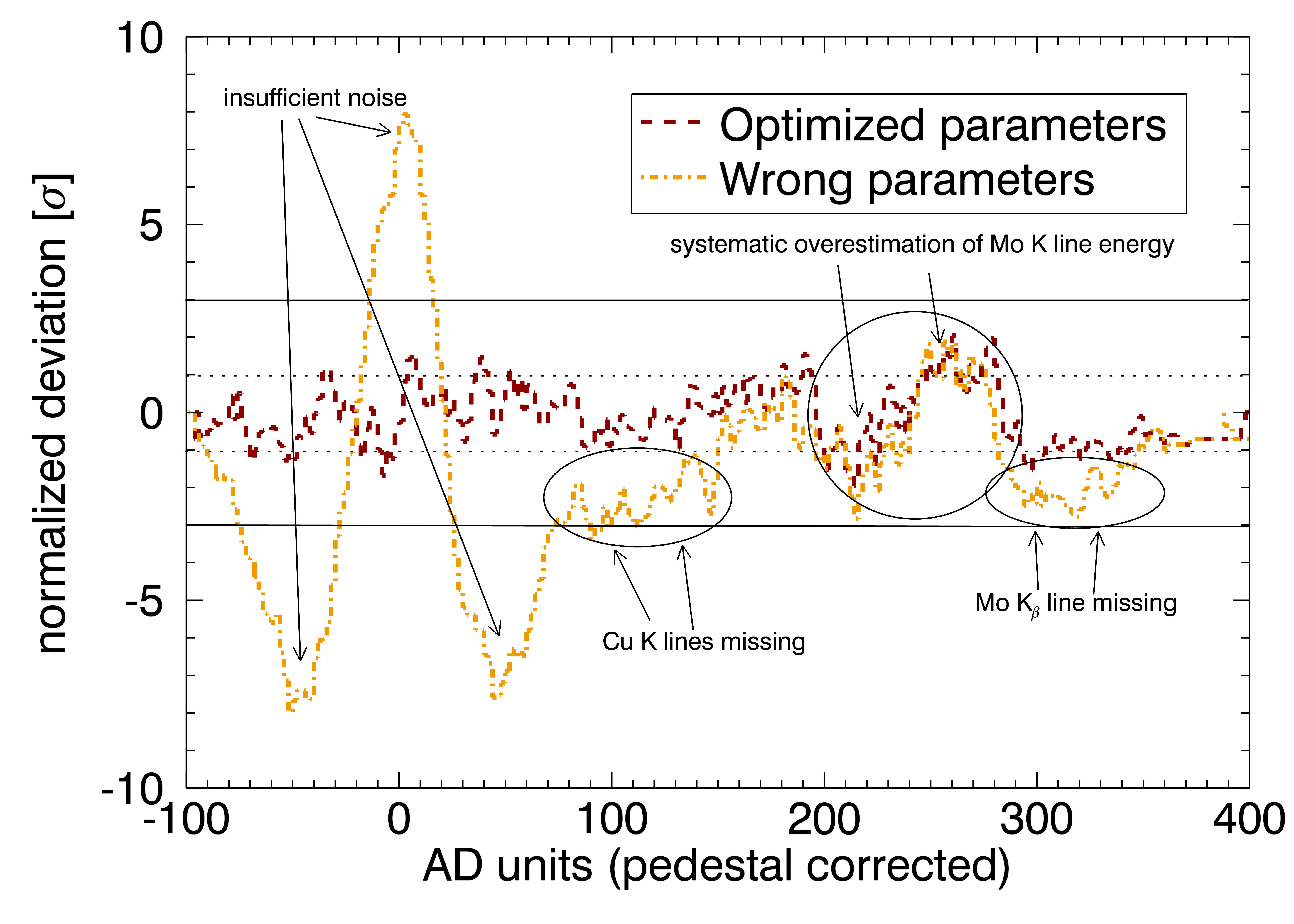}
  \centering
  \caption{Difference between the simulations and the measurement in units of the measurement error. The values show the expected statistical spread. Certain features are annotated to guide the eyes. For reference the horizontal lines indicate the 3 (solid) and 1 (dotted) sigma deviations. Systematic errors (e.g. effects not accounted for in the simulations) show correlated deviations from the otherwise random distribution.}
  \label{scatter}
\end{figure}

Figure \ref{scatter} shows the difference between simulation and measurement in units of $\sqrt{2}$ times the measurement error (square root of twice the number of counts in each bin\footnote{As there are two independent random processes, measurement and simulation.}). A perfect match with an infinite number of samples would show samples of a normal distribution with an rms (root-mean-square) value of 1.0 and a vanishing mean. As observed from figure \ref{scatter} there are some systematic deviations around the position of the Molybdenum fluorescence lines. It can be hypothesized that these deviations are caused by backscattering\footnote{One example of backscattered photons are photons that lose some energy by Compton scattering outside the sensitive volume, e.g. at the enclosure, chip tester box or air, and afterwards reach the sensitive volume.}, which is not included into HORUS. The simulations were performed 1000 times and mean and rms deviation were determined for each simulation. The average mean deviation is -0.099 $\pm$ 0.043 with an rms value of 1.127 $\pm$ 0.044. For comparison the corresponding values for the 'wrong' parameters are: -1.578 $\pm$ 0.032 and 3.052 $\pm$ 0.039.

As shown the chosen method is very sensitive to mis-tuned parameters and is therefore suitable for the validation and verification of the simulation package.

\section{Investigated performance parameters}

To investigate the single photon sensitivity of AGIPD a simple binary decision problem is used with the following two hypotheses: The null hypothesis (0), no x-ray in the sensitive volume, and the detection hypothesis (1), at least one x-ray in the sensitive volume.

The decision task is performed by thresholding the signal of a pixel using the threshold $Th$. This work will focus on two of the four possible outcomes ($P(1|1)$ and $P(1|0)$) which are explained in the following:

$P(1|1)$, correctly rejecting the null hypothesis: This is the probability to detect at least one photon when at least one photon is present in the sensitive volume. For the sake of simplicity a single event of this type will be called true positive and the probability true positive rate. Take note that it is normalized to photons in present in the sensitive volume, whether they are interacting in the sensitive volume or not. In order to renormalize to the number of incident photons it has to be multiplied by the quantum efficiency of the sensor which is shown in figure \ref{qe_window}.

$P(1|0)$, incorrectly rejecting the null hypothesis (error of the first kind): This is the probability to wrongly detect at least one photon when no photons are present in the sensitive volume.  For the sake of simplicity a single event of this type will be called false positive and the probability false positive rate. Take note that for matters of convenience only the expected number of false positives per frame of one megapixel will be shown, which is the false positive rate multiplied by the total number of pixels of the AGIPD.

Charge sharing and in-sensor scattering may cause an event not being detected in the pixel where its primary interaction took place. Therefore two additional performance parameters, which are not derived from the binary decision problem, are of interest: The expected number of pixels above threshold surrounding a pixel in which a primary interaction\footnote{At the investigated energy range possible primary interactions are (in order of decreasing likelihood): photoabsoption, coherent scattering and incoherent scattering. The probabilities of other interactions are many orders of magnitude lower.} took place (probability of misallocated events), and the additional number of events created in the local neighborhood (probability of multiple counts). 

Therefore the total detection efficiency (i.e. the probability to detect a photon anywhere in the local neighborhood) is the sum of the true positive rate (photons detected at the correct position) and the misallocation probability (photons detected at the wrong position).

\subsection{Analytic estimate of the false positive rate}

For Gaussian distributed random noise in a pixel, a reasonable approximation for the expected number of false positives per frame, in absence of photons, is:
\begin{equation}
	N_{false}= 0.5 \; N_{pix}\; \mathrm{erfc}\left(\frac{Th}{\sqrt{2} \; ENC}\right),  \label{analytic}
\end{equation}
where $N_{pix}$ is the number of detector pixels and $\mathrm{erfc}(x)$ the complementary error function. 

As equation \ref{analytic} only depends on the ENC (in electrons) we can use it to derive the required absolute threshold (in keV) , yielding $Th >  ENC / 58.3$ for the desired outcome of less than 1 false positive per frame\footnote{Therefore and ENC of 260 and 360 electrons requires a threshold of 4.46~keV and 6.17~keV, respectively.}.

It should be noted that this approximation is not valid in the presence of photons as misallocated events and multiple counts, which technically should be considered false positives, depend on the overall photon flux and energy. Additionally using the off-chip charge summing scheme invalidates the underlying approximations as discussed in section \ref{summing}.

\subsection{Definition of single photon sensitivity}

As mentioned in the introduction, for this study we call a system single photon sensitive when it is possible to find an operating point at which $\bar{P}(1|1) > 0.5$  and $P(1|0) < 10^{-6}$. 
Both  $\bar{P}(1|1)$ and $P(1|0)$ are functions of the systems noise and the threshold used for data evaluation. The threshold is restricted to the interval from $0.5 E_\gamma$ to $E_\gamma$, as lower thresholds result in significant multiple counting of events and higher thresholds have $\bar{P}(1|1) < 0.5$ by definition. This manuscript will present all threshold dependencies relative to the beam energy $E_\gamma$. For thresholds close to 1.0 $P(1|1)$ depends on the photon position within the pixel, therefore we will use $\bar{P}(1|1)$ in these cases, which is $P(1|1)$ averaged over all possible photon positions within the pixel.

The specific values used for these criteria are somewhat arbitrary, but have been proven to be useful in many experimental conditions. 

Failing to achieve single photon sensitivity at a specific energy does not render the detector incapable of imaging at this energy, it just means that the signal to noise ratio at this energy is worse than at energies where single photon sensitivity is achieved.

\subsection{Simulation method}

The simulation results for the false positive rate were derived a sample sizes of more than $10^{8}$ pixels without illuminating the sensor. Therefore the derivation of $P(1|0)$ is sensitive to approximately 1 false positive in 100 frames of 1MP each.

All other parameters were extracted from simulating more than $10^{7}$ neighborhoods of $5\times 5$ pixels separated from each other. Each neighborhood had a single photon of the given beam energy interacting in its sensitive volume. The interaction depth (Z coordinate) and type were sampled using the Monte Carlo method using the cross sections corresponding to the photon energy, the X and Y position within the pixel were sampled randomly. Scattered photons were tracked until they left the sensor volume.

The microbeam scan mentioned for 5~keV operation was simulated as mentioned above but the X and Y coordinates within the pixel were sampled randomly within the beam area (as opposed to the entire pixel area), which was scanned over the entire pixel surface.

\section{Results}

\begin{figure}[tb!]
	\centering
	\includegraphics[width=0.45\textwidth]{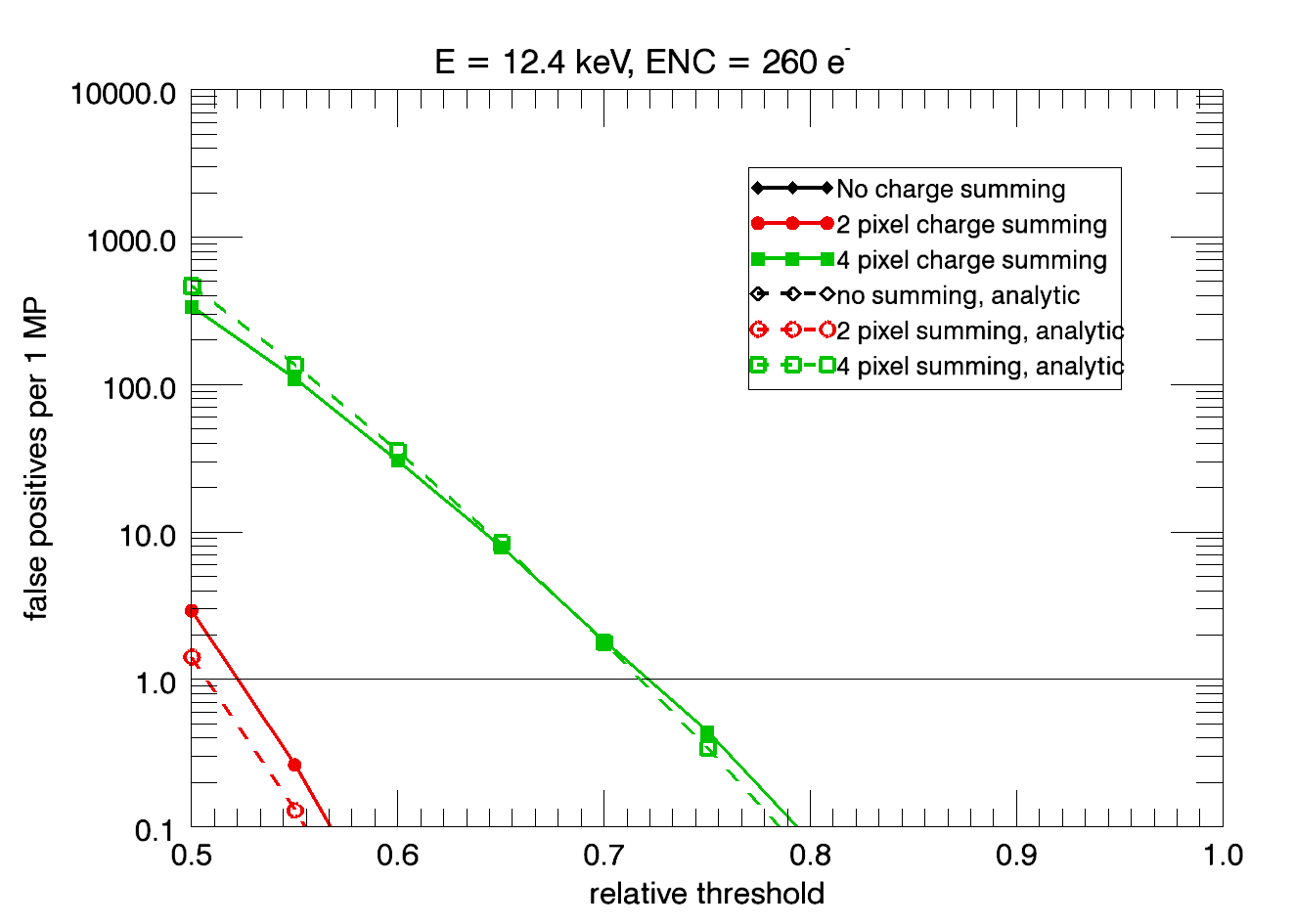}
	\includegraphics[width=0.45\textwidth]{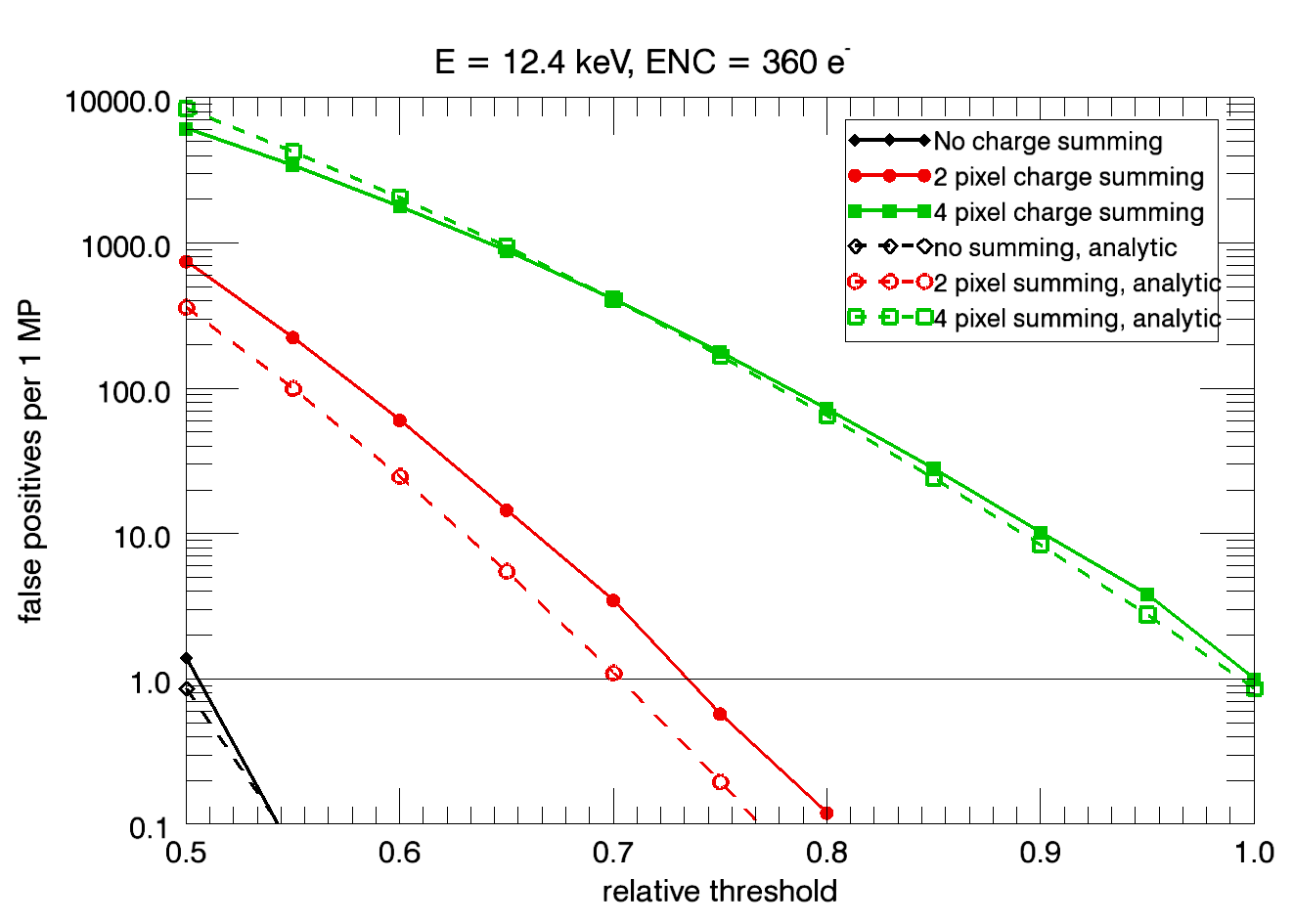} 	
	\caption{Number of false positives in a 1 MP detector for an ENC of 260 electrons (top) and 360 electrons (bottom) as a function of relative threshold and charge summing scheme. Values below 0.1 false positive per frame are not displayed. Without charge summing at an ENC of 260 electrons the false positive rate is consistently below 0.1. The dashed lines are calculated using equation \ref{analytic} and the approximation for charge summing mentioned in section \ref{summing}.}
	\label{12kev_cut}
\end{figure}

All results are normalized to the number of photon interactions in the sensitive layer or per MP in case of the false positives as indicated above. In order to account for the entrance window and limited quantum efficiency of the sensitive volume (renormalizing to the number of incident photons) the true positive rate has to be multiplied with the quantum efficiency $QE$ for the corresponding window (shown as a function of energy in figure \ref{qe_window}).

\subsection{Design energy: 12.4 keV}

As shown in figure \ref{12kev_cut}, less then one false positive per 1 MP can be achieved throughout the entire investigated noise range. It is not shown, but the true positive rate for low thresholds is close to the theoretical maximum value of approximately 97\%\footnote{The theoretical maximum value of the probability to detect a true positive is not unity, as there is about 3\% probability of a photon scattering and being absorbed in a different pixel at this energy.}. For larger thresholds the detection efficiency drops a little but stays above 50\%, similar to the behavior shown in upper image in figure \ref{8kev_cut}, which will be discussed in more detail later. Therefore single photon sensitivity at 12.4~keV beam energy is achieved. 

In most of the investigated noise range single photon sensitivity can also be obtained when using charge summing schemes. The details of the used summing algorithm and why 2 pixel summing produces more false positives than analytically expected and 4 pixel summing produces sometimes more and sometimes less is explained in section \ref{summing}. It should be noted that the employed charge summing is a data processing step which is done entirely offline, i.e. neither in the ASIC nor in the readout chain.

The impact of charge summing is detailed in the next paragraph using an example of 8~keV photons and low noise.

\subsection{Reduced beam energy: 8 keV}

\begin{figure}[tb!]
	\includegraphics[width=0.45\textwidth]{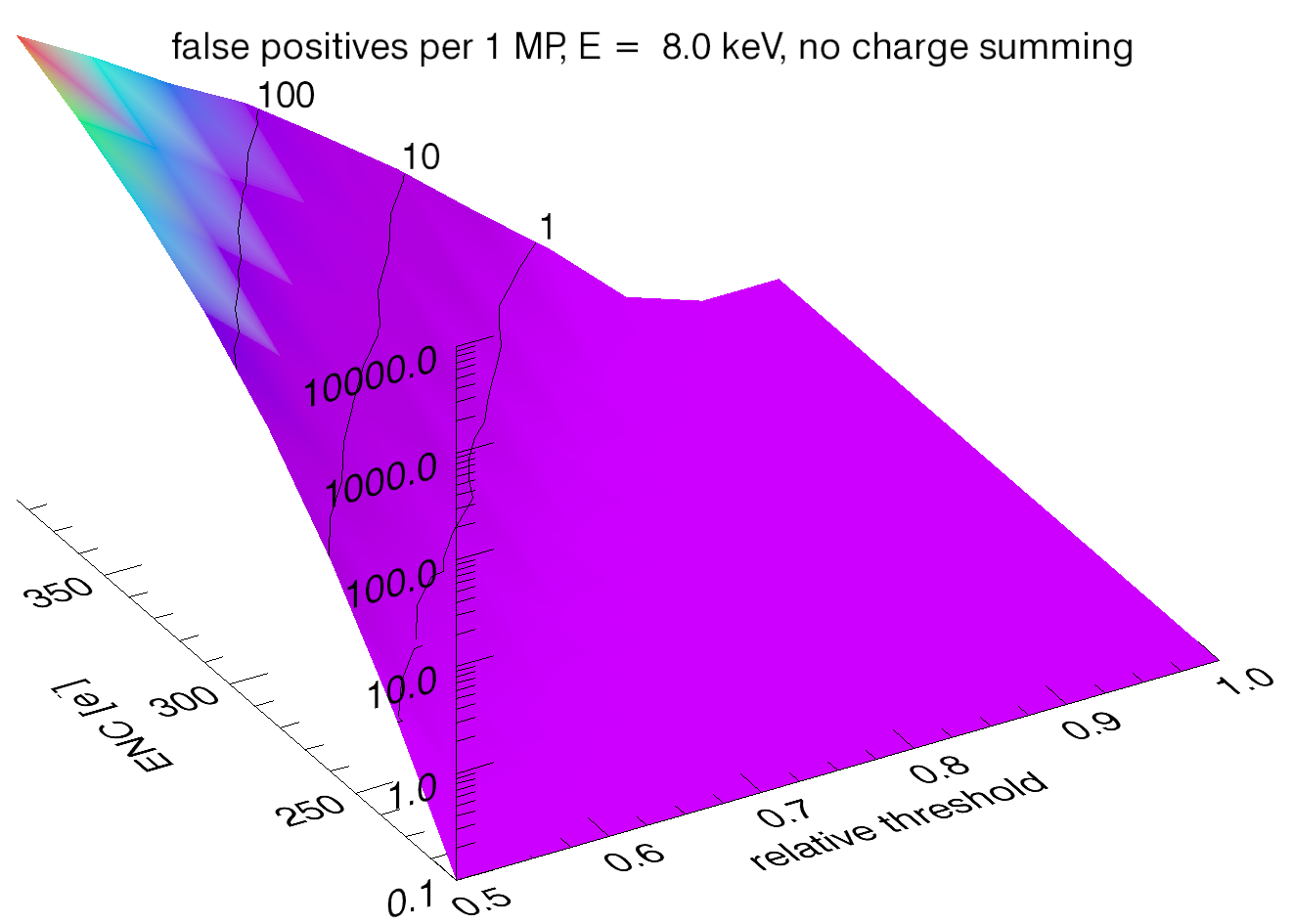}
  \centering
  \caption{False positives in a 1 MP detector as a function of noise and relative threshold.  Values below 0.1 false positive per frame are not displayed.}
  \label{8kev}
\end{figure}

\begin{figure}[tb!]
	\centering
	\includegraphics[width=0.45\textwidth]{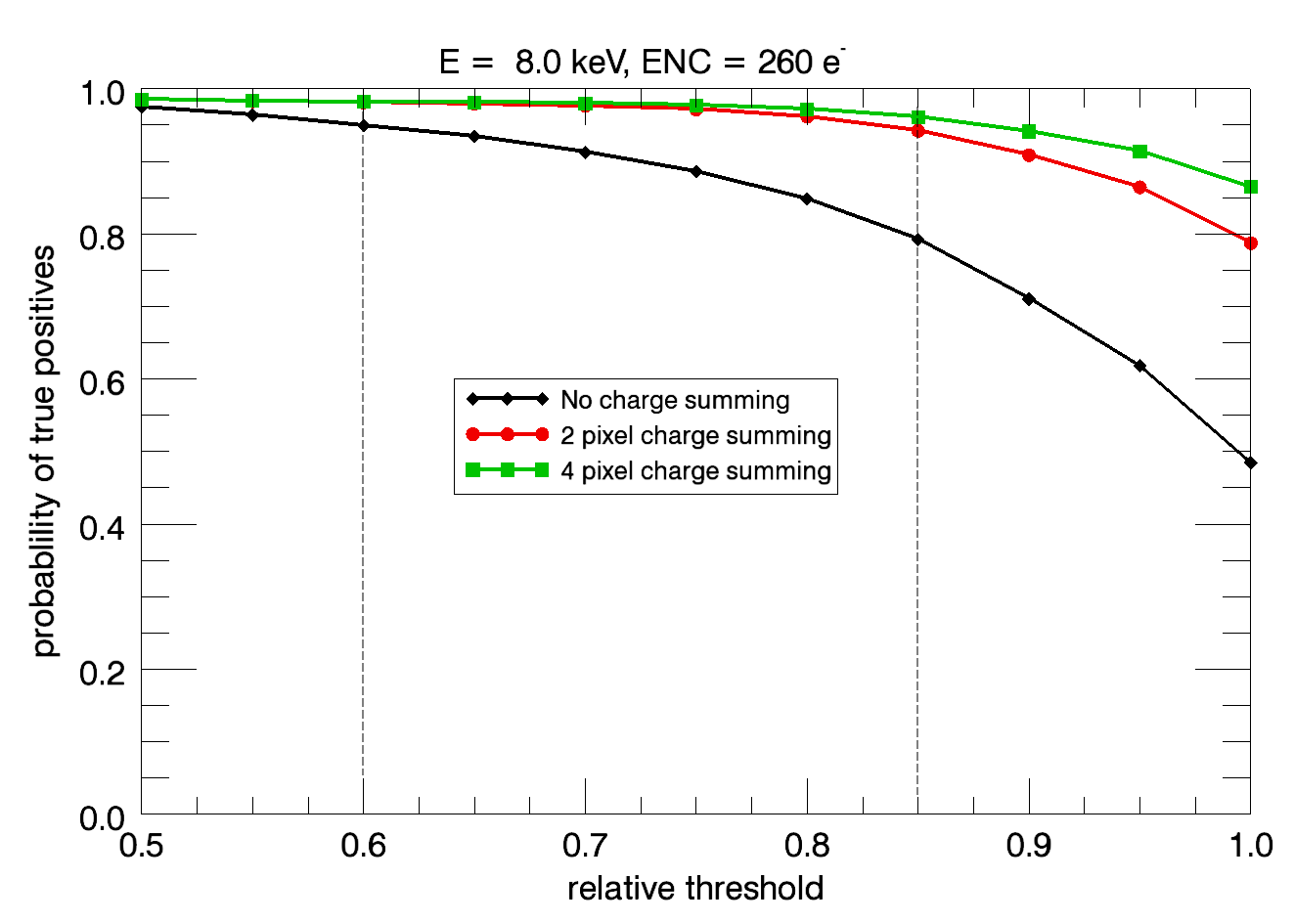}
	\includegraphics[width=0.45\textwidth]{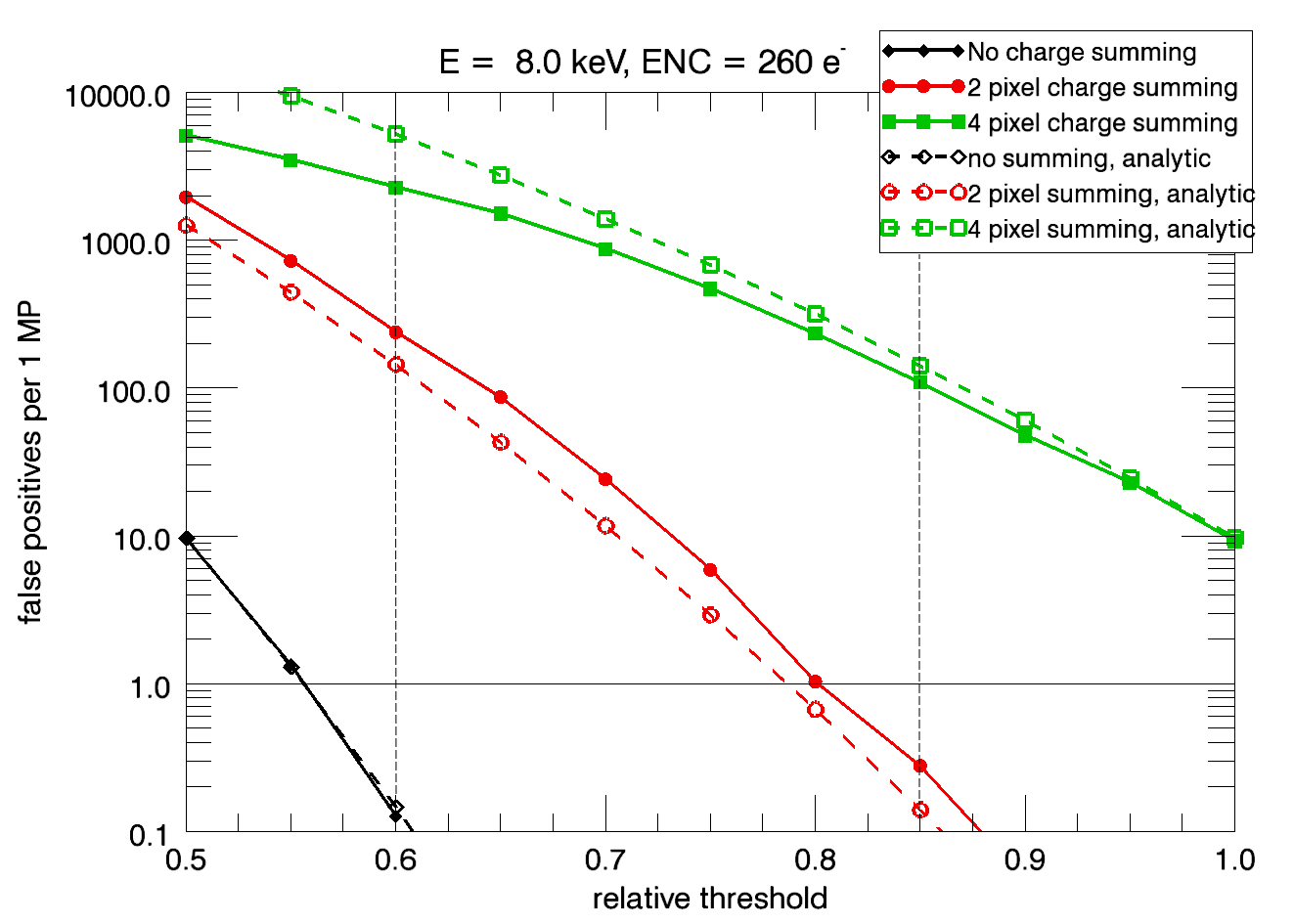} 	
	\caption{True and false positive rates in a 1 MP detector for a fixed ENC of 260 electrons as a function relative threshold and charge summing scheme. The thresholds needed to reduce the false positives to less then 1 per MP are indicated by the horizontal and vertical dashed lines. The dashed lines are calculated using equation \ref{analytic} and the approximation for charge summing mentioned in section \ref{summing}.}
	\label{8kev_cut}
\end{figure}

\begin{figure}[tb!]
	\centering
		\includegraphics[width=0.45\textwidth]{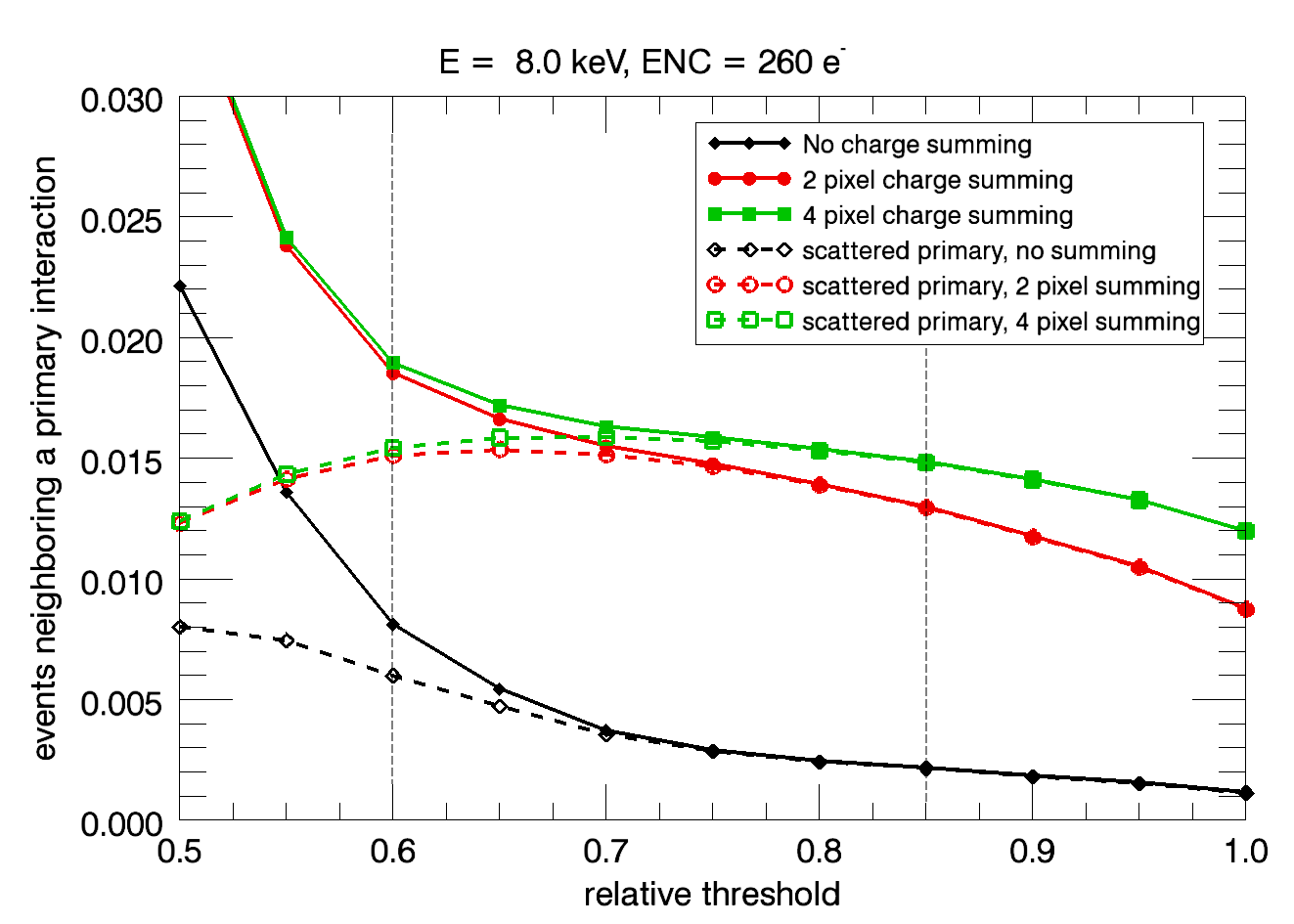}
		\includegraphics[width=0.45\textwidth]{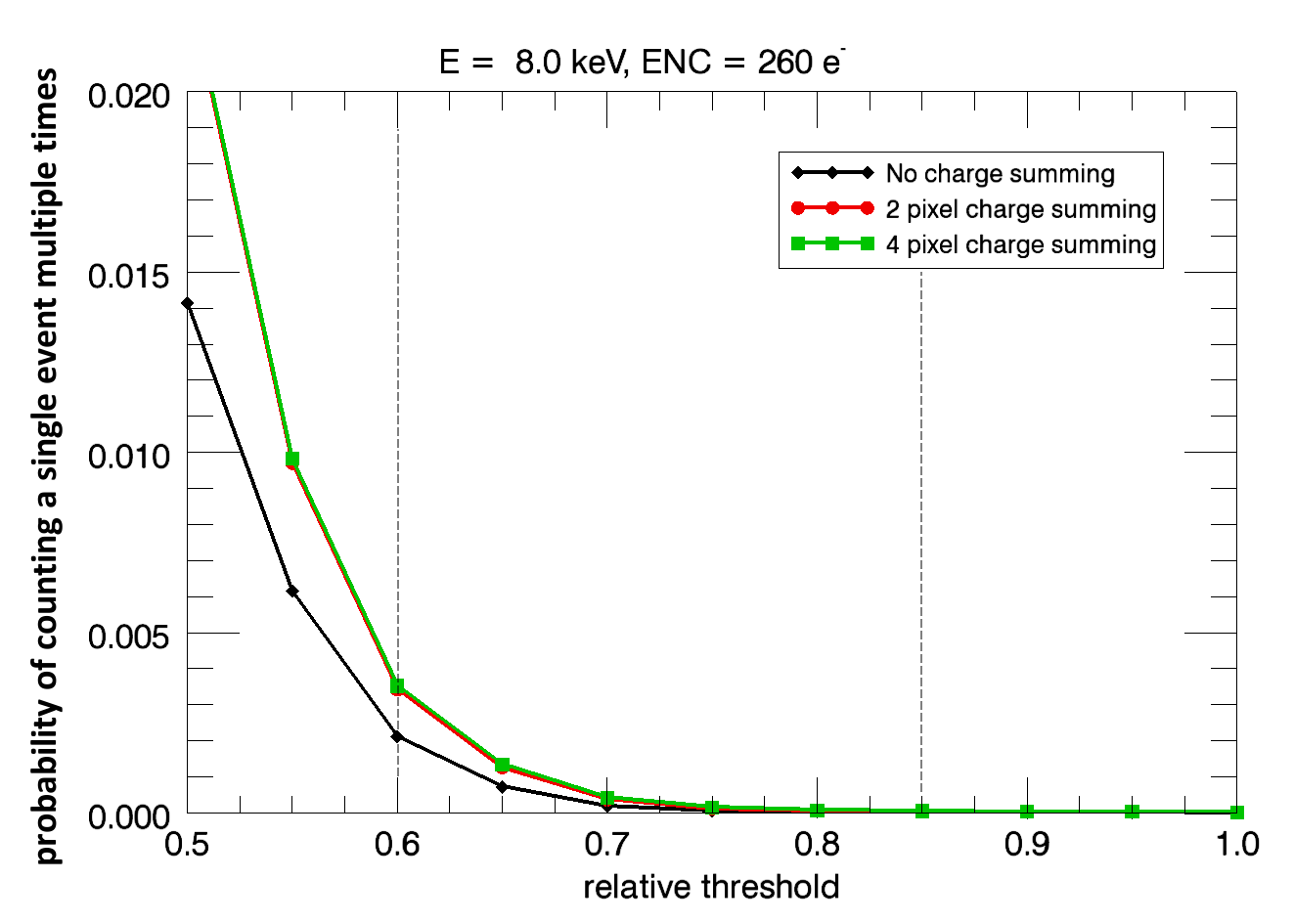} 	
	\caption{The solid lines of the upper image show the expected number of events above threshold in the local neighborhood excluding the central pixel. The number of events surrounding a central pixel that did not have a signal above threshold is shown by the dashed lines (this is e.g. the case when the primary photon scattered out of the sensitive volume of the central pixel). The lower image shows the average number of excess events a primary photon produces (i.e. the probability to count a photon multiple times). The thresholds needed to reduce the false positives to less then 1 per MP are indicated by the dashed lines.}
	\label{8kev_summing}
\end{figure}

Figure \ref{8kev} shows the situation for a beam energy of 8~keV and no charge summing. Single photon sensitivity can be achieved over the whole investigated noise interval (not shown). Charge summing, however, can only be applied at low noises if single photon sensitivity shall be kept (see figure \ref{8kev_cut}).

The situation for high noise levels is very similar to the situation for low noise levels and 5~keV beam energy, which is detailed in the next paragraph. The situation for low noise is detailed here.

Figure \ref{8kev_cut} shows the true and false positive rate for a fixed ENC of 260 electrons. If $P(1|0) < 10^{-6}$ is required, the relative threshold has to be increased to approximately 0.6, when no charge sharing scheme is used and to approximately 0.85 when the 2 pixel summing scheme is used\footnote{In the simulations the relative threshold was increased in steps of 0.05 and thresholds of 0.55 and 0.8 produced slightly more than 1 false positive per MP.}. At these thresholds the true positive rate is approximately 95\% for both.

The benefit of employing charge summing is shown in figure \ref{8kev_summing}.The solid lines of the upper image show the expected number of events above threshold in the local neighborhood excluding the central pixel. The number of events surrounding a central pixel that did not have a signal above threshold is shown by the dashed lines (for example, when the primary photon is scattered out of the sensitive volume of the central pixel). The lower image shows the average number of excess events a primary photon produces (i.e. the probability to count a photon multiple times, which is the difference between the solid and dashed lines in the upper image).

For the thresholds mentioned above, using 2 pixel charge summing produces the same total detection probability, but 0.7\% points more events are displaced (difference of the dashed lines in the upper plot of figure \ref{8kev_summing}).

Not using charge summing results in 0.2\% additional events in the local neighborhood, i.e. every 500th event is counted as two events in adjacent pixels. When using 2 pixel charge summing only approximately $5 \times 10^{-5}$ events (1 every 20'000) are counted as two events. This is a reduction by a factor of 40.

At an ENC of 360 electrons the results are similar to the results obtained for 5~keV energy and an ENC of 260 electrons, which are discussed below.

\subsection{Extended energy range: 5 keV}

\begin{figure}[tb!]
	\includegraphics[width=0.45\textwidth]{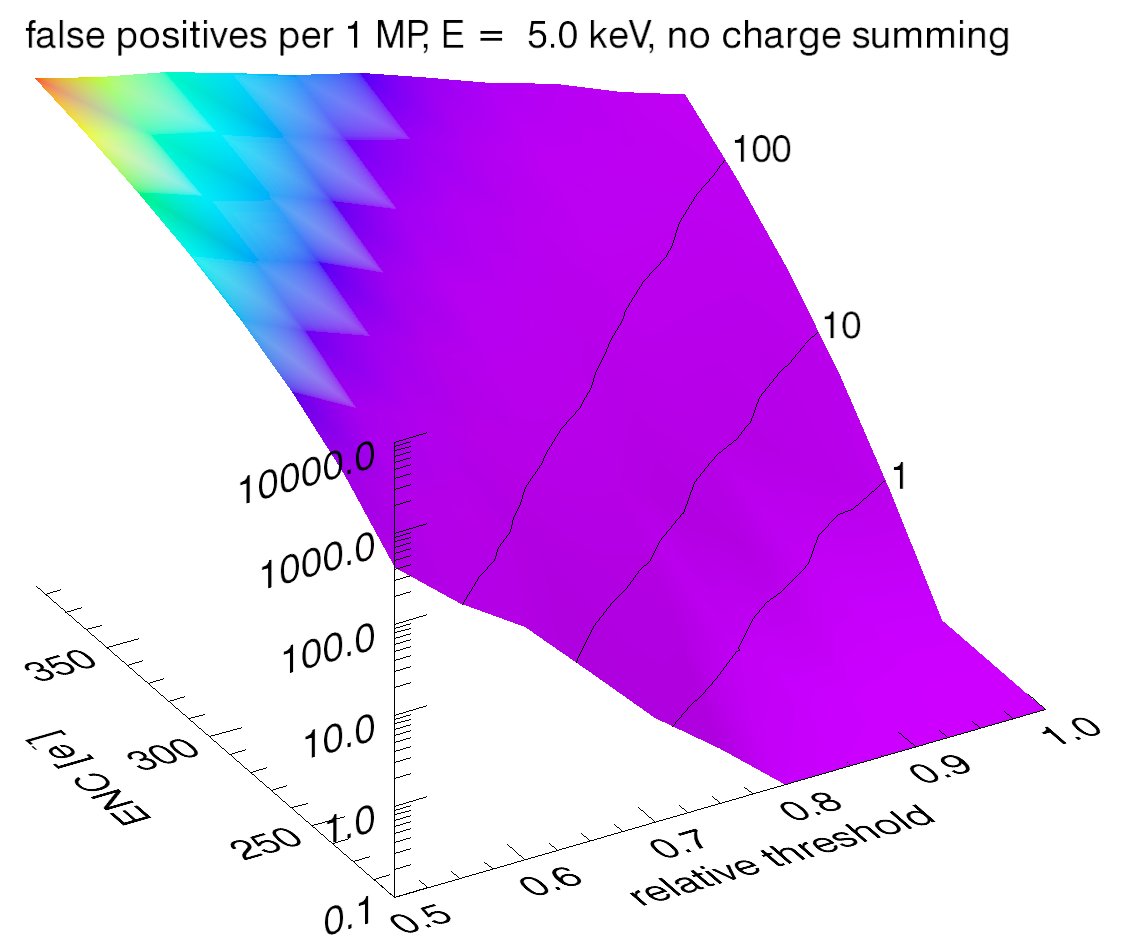}
  \centering
  \caption{False positives in a 1 MP detector as a function of noise and relative threshold . Values below 0.1 false positive per frame are not displayed.}
  \label{5kev}
\end{figure}

\begin{figure}[tb!]
	\centering
		\includegraphics[width=0.45\textwidth]{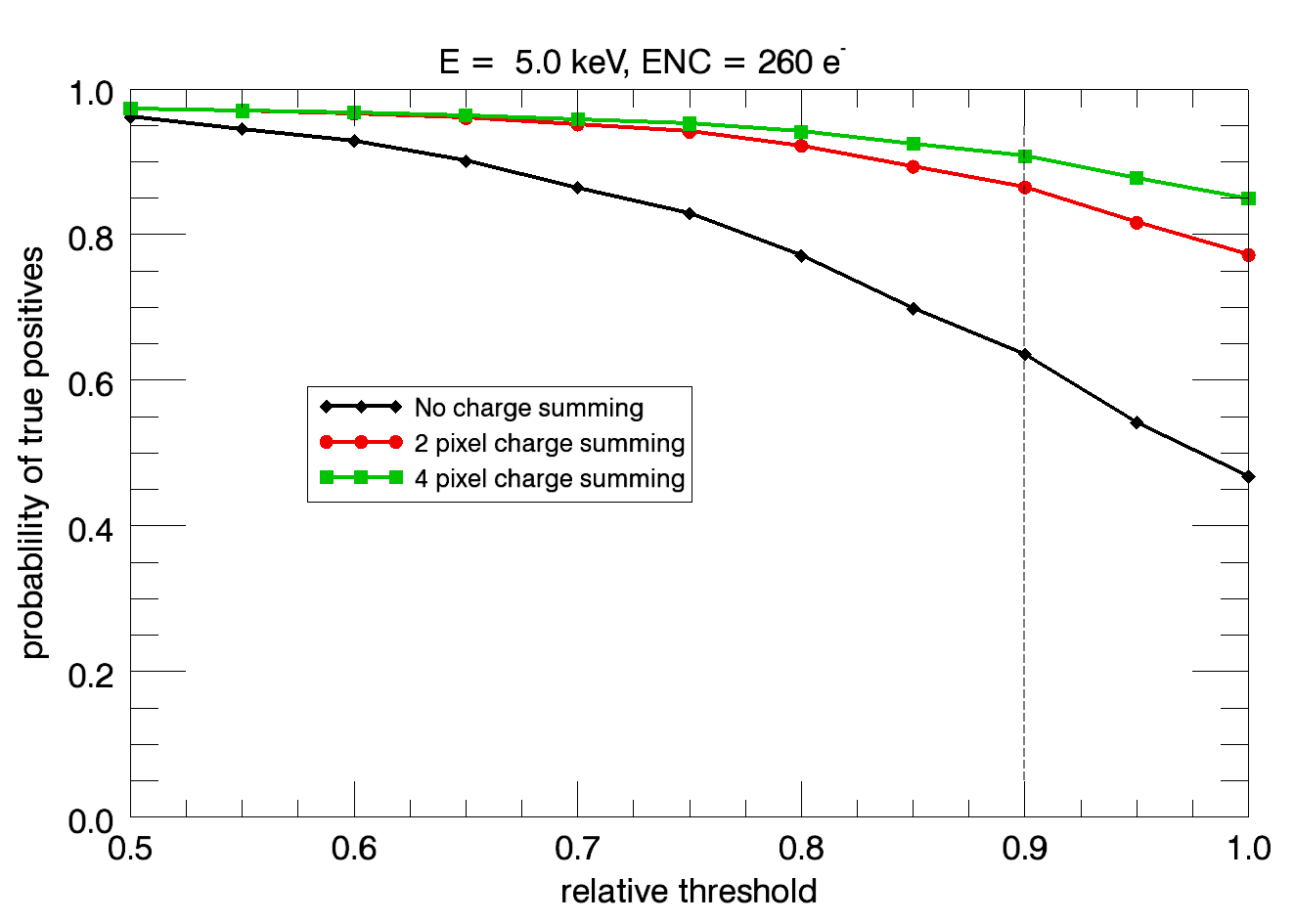}
		\includegraphics[width=0.45\textwidth]{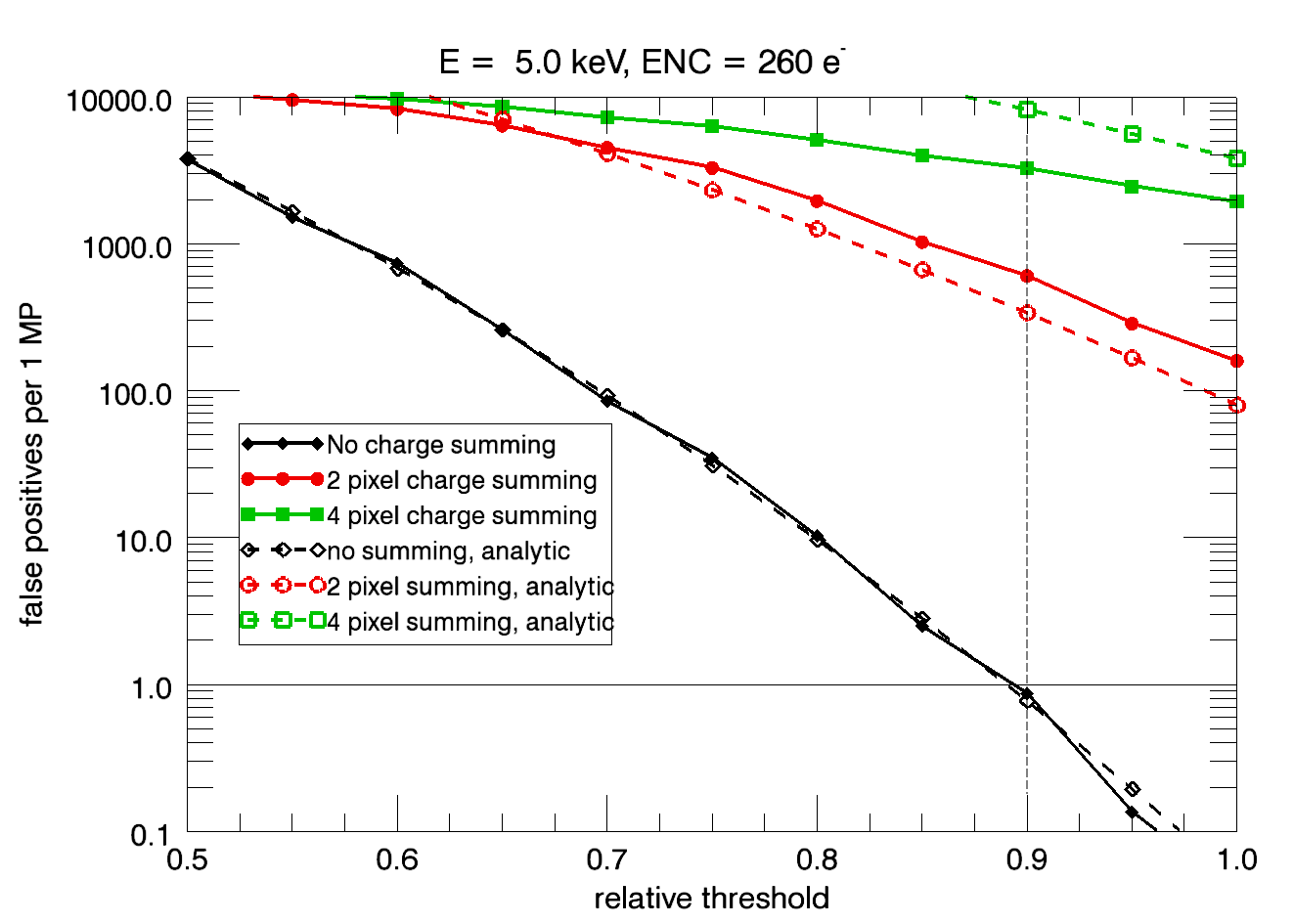} 	
	\caption{True and false positive rates in a 1 MP detector for a fixed ENC of 260 electrons as a function relative threshold and charge summing scheme. The threshold needed to reduce the false positives to less then 1 per MP is indicated by the horizontal and vertical dashed lines. The dashed lines are calculated using equation \ref{analytic} and the approximation for charge summing mentioned in section \ref{summing}.}
	\label{5kev_cut}
\end{figure}

\begin{figure}[tb!]
	\begin{tabular*}{0.45\textwidth}{@{}c@{}@{}c@{}} 
			\begin{tabular*}{0.40\textwidth}{@{}c@{}} 
				\includegraphics[width=0.40\textwidth]{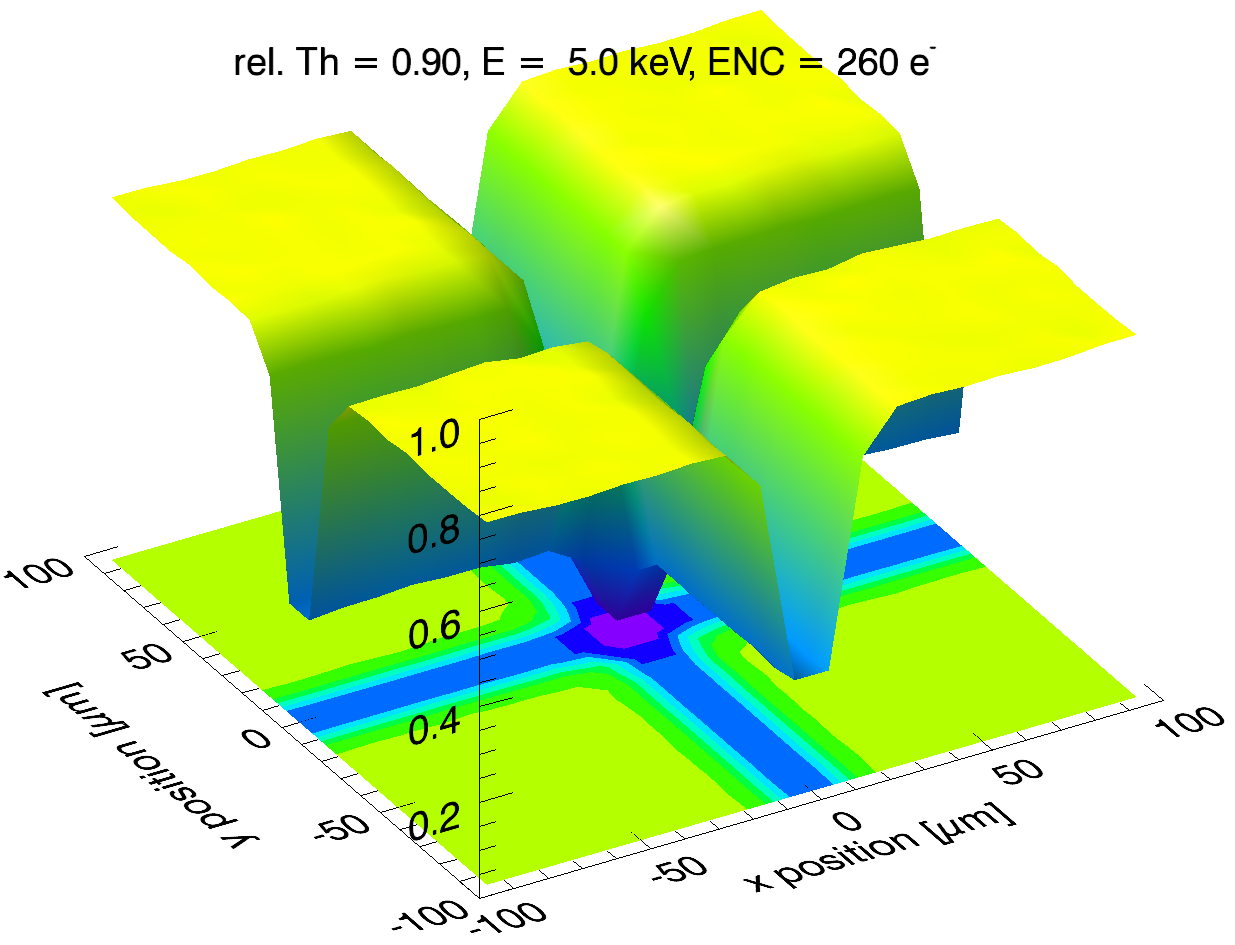}
			\end{tabular*} &
			\begin{tabular*}{0.05\textwidth}{@{}c@{}}
				\includegraphics[width=0.05\textwidth]{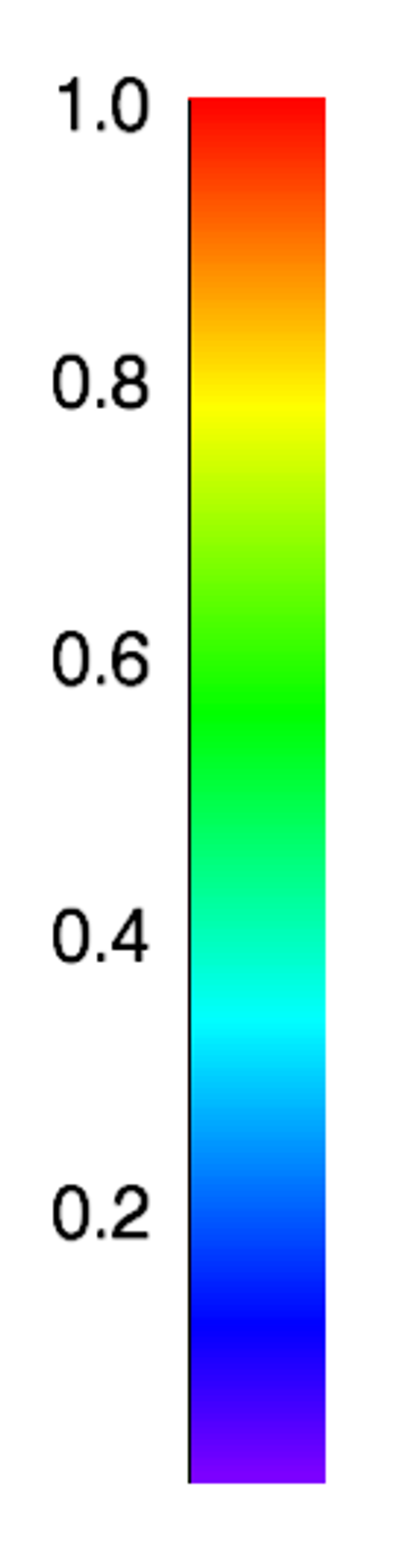} 
			\end{tabular*}
	\end{tabular*}
  \centering
  \caption{Microbeam scan using a beam size of (10~$\upmu$m)$^2$ and a beam energy of 5~keV. The origin of the coordinate system is centered on the pixel corner.}
  \label{5kev_microbeam}
\end{figure}

For a beam energy of 5~keV single photon sensitivity can be achieved in a low noise scenario, as shown in figure \ref{5kev} and \ref{5kev_cut}. At an ENC of 260 electrons a relative threshold of 0.9 has to be selected to achieve single photon sensitivity, which in turn leads to an average true positive rate of approximately 63\% (see figure \ref{5kev_cut}). 

At high relative thresholds $P(1|1)$ is no longer uniform over the pixel area, but decreases strongly towards the edges and corners of a pixel as a result of the extend of the charge cloud generated by the individual x-rays. The simulation of a microbeam scan visualizing this effect is shown in figure \ref{5kev_microbeam}. While in the central region of a pixel $P(1|1)$ is approximately 80\%, it decreases sharply towards a pixels edge, vanishing almost completly at the corner. The microbeam scan used a square beam with a size of (10~$\upmu$m)$^2$ and an energy of 5~keV. The origin of the coordinate system is centered on the pixel corner, and any event registered by one of the four pixels sharing this corner is counted.

\subsection{Low energy operation: 3 keV}

\begin{figure}[tb!]
	\centering
	\includegraphics[width=0.45\textwidth]{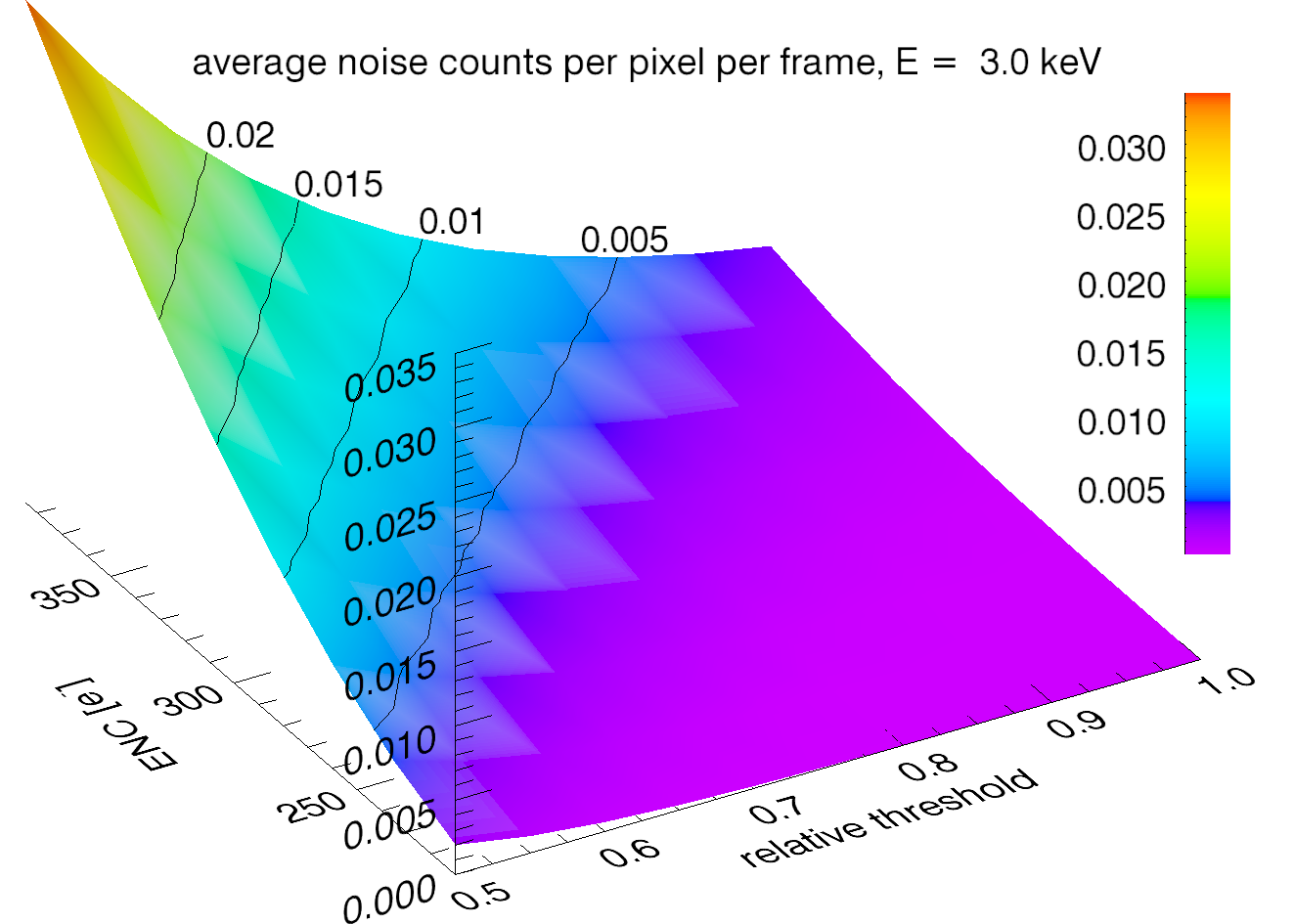}
	\caption{Average number of photon 'hits' due to noise per pixel per frame as a function of noise and relative threshold.}
	\label{3kev_avg}
\end{figure}

Single photon sensitivity, as defined in this work, cannot be achieved at 3~keV beam energy. The average count rate of false 'hits' is below 0.035 per pixel per frame (see figure \ref{3kev_avg}, depending on noise and threshold a reduction by a factor $>10$ can be achieved), but significantly above the required 10$^{-6}$. Due to the Gaussian nature of the noise no more than 1 or 2 'false' photons will be detected in an individual pixel, depending on ENC and threshold. 

Even though single photon sensitivity is not given, imaging at this energy is still possible. When the relative threshold is chosen below 0.7, the average probability to detect an interacting photon is above 70\%\footnote{The probability to detect impinging single photons has to be multiplied by the QE as given in Figure \ref{qe_window}}. As all integrating detectors the AGIPD will be limited by Poisson noise for signals significantly above the noise floor.

\section{Summary}

\begin{table*}
\centering	
\begin{tabular}{c|c|c|c|c}
				& \textbf{threshold}	& \textbf{true positive rate [\%]}	& \textbf{misallocated events [\%]}	& \textbf{multiple counts [\%]} \\
\hline
\textbf{12.4 keV}		& 0.50/0.55 			& 96.7/95.5					& 1.4/1.5 					& 1.20/0.78  \\
\textbf{2 pixel summing}	& 0.55/0.75 			& 97.1/96.1 					& 1.7/2.1 					& 0.57/0.29  \\
\textbf{4 pixel summing}	& 0.75/1.00 			& 97.0/85.7 					& 1.8/1.7 					& 0.30/0.14 \\
\hline
\textbf{8 keV}		& 0.60/0.85 			& 94.9/80.4 					& 0.59/0.23 					& 0.2/0.005 \\ 
\textbf{2 pixel summing}	& 0.85/- 			& 94.2/- 					& 1.29/- 					& 0.005/- \\ 
\hline
\textbf{5 keV}		& 0.90/- 			& 63.5/- 					& 0.064/- 					& 0.0035/- \\
\end{tabular}
\caption{Detector performance results presented in this study. Results are presented for the exemplary 260 and 360 electrons noise, respectively. Shown results are for the indicated relative thresholds, which are the minimum thresholds required to achieve single photon sensitivity. For configurations of beam energy and charge summing scheme not presented here single photon sensitivity could not be achieved. It should be noted that although the threshold is increased when employing charge summing schemes, the true positive rate stays approximately the same although in total more events are detected. Furthermore it is observed that charge summing reduces the probability of counting an event multiple times at the expense of a higher probability to have an event misallocated.}
\label{summary}
\end{table*}

This work investigated the single photon sensitivity of the AGIPD system. Single photon sensitivity was defined as the possibility to select a threshold such that the expected rate for false positives is below one per frame ($P(1|0) < 10^{-6}$), while simultaneously having an average true positive rate of more than 50\% ($\bar{P}(1|1) > 0.5$).

Simulations were performed using the HORUS detector simulation toolkit, assuming the noise of the final AGIPD to lie in the interval between 200 and 400 electrons.

It was shown that AGIPD is single photon sensitive throughout the investigated noise interval down to 8~keV beam energy. Should the final noise be at the lower end of the possible range (e.g. 250 electrons) single photon sensitivity can also be achieved at 5~keV beam energy. A short summary of the results obtained for 260 and 360 electrons noise is shown in table \ref{summary}.

It was shown that charge summing schemes are beneficial when the noise is sufficiently low. The total detection rate of events increased (although some of the events were reconstructed in neighboring pixels) and the possibility to count one event twice in adjacent pixels is reduced by a factor of up to 40. Charge summing can be used at the design energy of 12.4~keV in the entire investigated noise regime, and at 8~keV if the noise is low (e.g. 250 electrons).

The low energy performance of AGIPD was explored, finding a noise floor below 0.035 'hits' per pixel per frame at 3~keV beam energy. Even though single photon sensitivity, as defined in this work, is not given, imaging at this energy is still possible, allowing Poisson noise limited performance for signals significantly above the noise floor.

\end{document}